# ОБ ОДНОМ СПЕЦИАЛЬНОМ СЛУЧАЕ ОБОБЩЕННОЙ ПРОБЛЕМЫ ОКРУЖЕНИЙ

В. Г. Найденко, Ю. Л. Орлович

For a given finite class of finite graphs $\mathcal{H}$, a graph $G$ is called a *realization* of $\mathcal{H}$ if the neighbourhood of its any vertex induces the subgraph isomorphic to a graph of $\mathcal{H}$. We consider the following problem known as the *Generalized Neighbourhood Problem (GNP):* given a finite class of finite graphs $\mathcal{H}$, does there exist a non-empty graph $G$ that is a realization of $\mathcal{H}$? In fact, there are two modifications of that problem, namely the *finite* (the existence of a finite realization is required) and *infinite* one (the realization is required to be infinite). In this paper we show that GNP and its modifications for all finite classes $\mathcal{H}$ of finite graphs are reduced to the same problems with an additional restriction on $\mathcal{H}$. Namely, the orders of any two graphs of $\mathcal{H}$ are equal and every graph of $\mathcal{H}$ has exactly $s$ $(s \geqslant 1)$ dominating vertices.

## Введение

Проблемы существования и единственности графа $G$ с заданным значением некоторого инварианта $I(G)$ занимают в теории графов видное место. К ним относится, в частности, известная проблема окружений [6, 28]: существует ли связный граф $G$, окружение каждой вершины которого порождает подграф, изоморфный данному конечному графу $H$, и если существует, то единственный ли с точностью до изоморфизма?

Проблема окружений служит классическим примером задачи восстановления [6] и решена для различных графов $H$ [1–3, 8–10, 12–13, 15–21, 23, 24, 26, 27]. Она рассматривалась в связи с вопросами теории автоматов, теории групп, теории конечных геометрий и в связи с приложениями в вычислительной технике.

В дальнейшем под *графом* $G = (VG, EG)$ понимается локально конечный неориентированный граф без петель и кратных ребер с не более чем счетным множеством вершин $VG$ и непустым множеством ребер $EG$.

Все понятия, не определенные в статье, можно найти в книгах [5–7, 11].

Число $|VG|$ вершин конечного графа $G$ называется его *порядком* и обозначается через $|G|$.

*Классом* графов называется любое множество конечных графов, различаемых с точностью до изоморфизма.

Граф называется *нетривиальным,* если либо он связен, либо среди его связных компонент нет изоморфных.

Множество всех вершин графа $G$, смежных с некоторой вершиной $v$, называется *окружением вершины* $v$ и обозначается через $N(v, G)$. Число $\deg_G v = |N(v, G)|$ – *степень вершины* $v$. Если граф $G$ зафиксирован, то $N(v, G)$ обозначим просто через $N(v)$.

Если $U \subseteq VG$, то $G(U)$ – подграф графа $G$, порожденный множеством вершин $U$.

Положим $N(G) = \{\langle G(N(v)) : v \in VG\rangle\}$, где $\langle \mathcal{F}\rangle$ – максимальное по включению множество попарно неизоморфных графов семейства $\mathcal{F}$.

Для некоторых графов мы используем стандартные обозначения: $K_n$, $O_n$, $C_n$, $P_n$ – соответственно полный граф (клика), пустой граф, простой цикл и простая цепь с $n$ вершинами, $K_{m,n}$ – полный двудольный граф с долями мощности $m$ и $n$. Кроме того, $\overline{G}$ обозначает граф, дополнительный к $G$, $L(G)$ – реберный граф, а $G^k$ – $k$-ю степень графа $G$.

Пусть $\mathcal{H}$ – конечный класс графов. Возникают следующие две задачи [6, 28].

1. *Существование.* При каком условии $N(G) = \mathcal{H}$ для некоторого графа $G$? Другими словами, когда существует какой-либо граф $G$, реализующий $\mathcal{H}$?

2. *Единственность.* Пусть $N(G_1) = N(G_2) = \mathcal{H}$ для некоторых графов $G_1$ и $G_2$. При каком условии отсюда вытекает изоморфизм $G_1 \cong G_2$?

Проблемы существования и единственности графа $G$ с заданным значением инварианта $N(G)$ называют *обобщенной проблемой окружений.* Граф $G$, удовлетворяющий условию $N(G) = \mathcal{H}$, называется *реализацией* класса $\mathcal{H}$ или *локально* $\mathcal{H}$*-графом*, а класс $\mathcal{H}$ – *реализуемым* классом графов. В частности, если все графы класса $\mathcal{H}$ изоморфны одному графу $H$, то локально $\mathcal{H}$-граф мы называем *локально* $H$*-графом* или просто *реализацией* графа $H$, а $H$ – *реализуемым* графом.

На рис. 1 в качестве иллюстрации приведен пример построения реализации класса $\{P_5, C_5\}$ на основе произвольного плоского кубического графа.

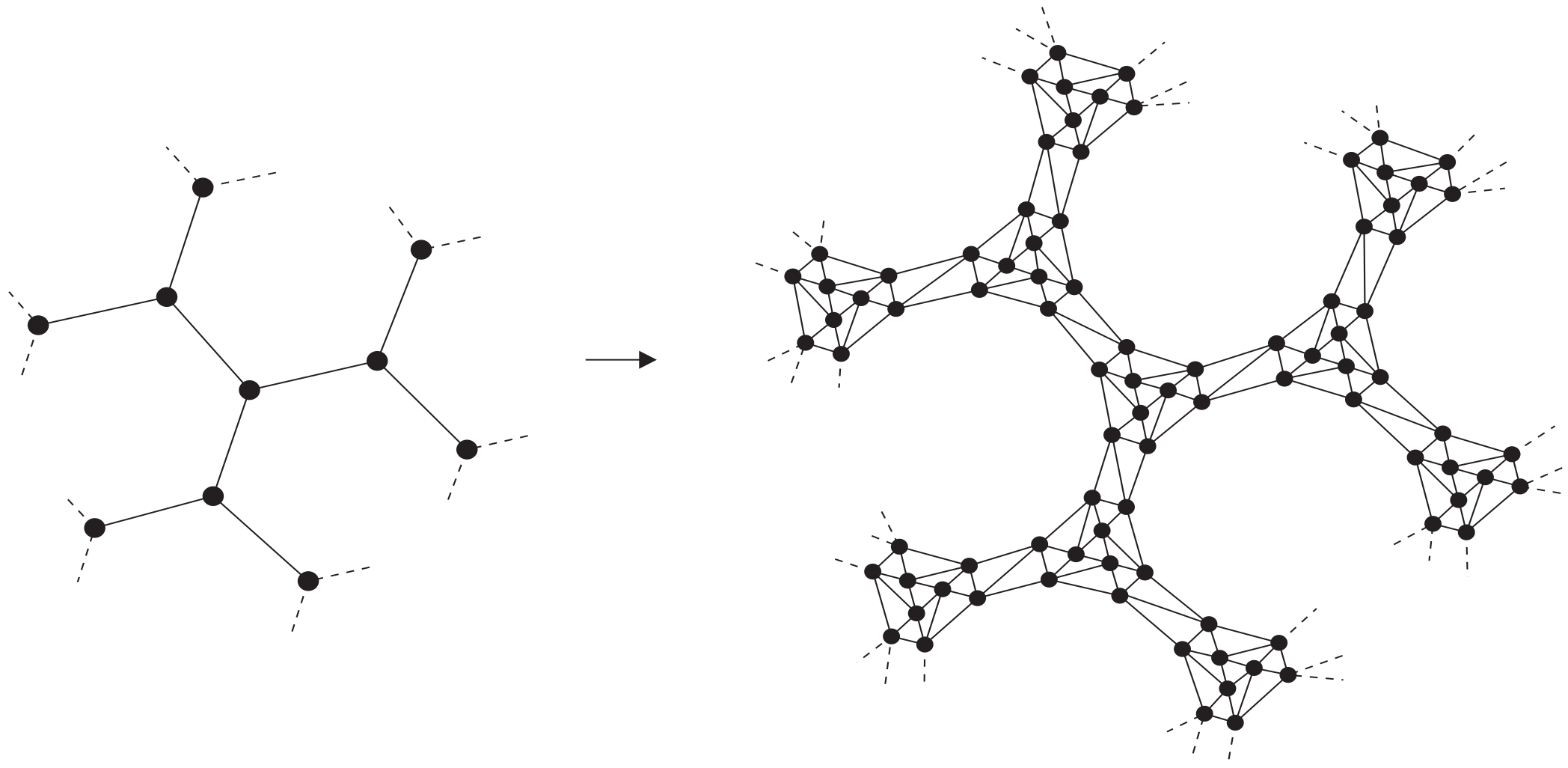

Рис. 1. Фрагмент построения реализации класса $\{P_5, C_5\}$.

В [1] указаны простые примеры классов $\mathcal{H}$, в частности, состоящих из одного графа, для которых существуют только бесконечные локально $\mathcal{H}$-графы. Поэтому в дальнейшем естественно различать конечные и бесконечные реализации, т.е. изучать множество $\{G : N(G) = \mathcal{H}\}$ при соответствующих глобальных ограничениях на граф $G$. В этом случае содержательны две следующие задачи.

3. *Проблема* $f$*-реализуемости.* По заданному конечному классу графов $\mathcal{H}$, выяснить, существует ли конечный граф $G$, реализующий $\mathcal{H}$?

4. *Проблема* $i$*-реализуемости.* По заданному конечному классу графов $\mathcal{H}$, выяснить, существует ли нетривиальный бесконечный граф $G$, реализующий $\mathcal{H}$?

Если при этом класс $\mathcal{H}$ состоит из графов, изоморфных некоторому графу $H$, то получаем соответственно *специальные* проблемы $f$-реализуемости и $i$-реализуемости. Интересно отметить, что, как было показано В. К. Булитко [3, 4], в классе всех конечных графов специальная проблема $i$-реализуемости алгоритмически неразрешима, т.е. справедлива

Теорема 1. *Невозможен алгоритм, распознающий по произвольному конечному графу* $H$ *существует ли связный бесконечный граф* $G$*, реализующий* $H$.

Доказательство теоремы 1 состоит в сведении проблемы сильной разрешимости конечного набора домино на плоскости, которая является модификацией алгоритмически неразрешимой “угловой” проблемы домино [11, 22] к специальной проблеме $i$-реализуемости. А именно по произвольному конечному набору $\mathcal{D}$ домино строится такой конечный граф $H_{\mathcal{D}}$, для которого реализация связным бесконечным графом $G$ существует тогда и только тогда, когда $\mathcal{D}$ сильно разрешим на плоскости. Отметим, что утверждение теоремы 1 справедливо также в ряде других случаев, в частности, когда $H$ – произвольный несвязный, связный, плоский или регулярный конечный граф [3, 4].

В работе [4] сведением проблемы разрешимости конечного набора домино на торе к проблеме $f$-реализуемости, установлена

Теорема 2. *Невозможен алгоритм, распознающий по произвольному конечному классу графов $\mathcal{H}$ существует ли конечный граф $G$, реализующий $\mathcal{H}$.*

Вопрос о том, разрешима ли в общем случае специальная проблема $f$-реализуемости, остается открытым (см., например, [14, 23]). В то же время в литературе известны нетривиальные примеры подклассов класса всех конечных графов, где специальные проблемы 3 и 4 алгоритмически разрешимы. Например, класс деревьев [2, 16] или более широкий класс графов, у которых каждый блок – полный граф [2, 13].

Заметим, что при наложении определенных мощностных ограничений на блоки последний класс будут образововать "плотные" графы [19, 23]. Класс $\mathcal{D}$ графов будем называть *плотным*, а сами графы $G \in \mathcal{D}$ – *плотными*, если выполняются следующие условия:

а) существует такая функция $\mathcal{E} : \mathbb{N} \to \mathbb{N}$, что $\lim\limits_{n\to\infty} \mathcal{E}(n)/n^2 = c$, где $c$ – константа, $c > 0$;

б) число ребер каждого графа $G \in \mathcal{D}$ порядка $n$ не меньше $\mathcal{E}(n)$.

Для многих плотных графов часто удается решить не только вопрос о существовании какой-либо реализации, но и получить полное описание всевозможных (в том числе и бесконечных) реализаций. Например, в [26] доказано, что если натуральные числа $t$ и $r$ связаны неравенством

$$3r < 3t + \sqrt{8(t-1)} + 4,$$

то $t$-регулярный граф порядка $r$ реализуем тогда и только тогда, когда $r - t$ делит $r$. В этом случае реализация – полный многодольный граф с $1 + r/(r-t)$ долями порядка $r - t$. Заметим, что в [12] получено описание локально $t$-регулярных графов порядка $r$ с более слабым условием на числа $t$ и $r$, а именно $t \geqslant r - \sqrt{r} - 1/2$.

Другие условия реализуемости плотных графов можно найти в [15, 17–19, 23, 27]. Так в [19] классифицированы реализуемые графы порядка $r$, дополнительные к несвязным графам с равномощными компонентами порядка $p$. Оказалось, что все они исчерпываются дополнением дизъюнктного объединения $r/p$ полных графов $K_p$. Заметим, что этот результат не перекрывается классификацией локально $t$-регулярных графов, полученной в [12, 26], ибо в общем случае условие равномощности компонент не влечет за собой регулярности дополнительного графа.

В [27] приведена характеризация локально $\overline{P}_3$-графов как реберных графов от графов, получающихся из кубических однократным подразбиением ребер, и найдены все локально $\overline{P}_n$-графы для $n \in \{1, 2, 4\}$ и $n \geqslant 5$. Там же выявлено описание локально $\overline{C}_n$-графов.

Классификация локально $C_n^k$-графов и локально $P_n^k$-графов при $2 \leqslant k \leqslant n-1$ получена в работах [17] и [18] соответственно. Установлено, что $C_{2k+2}^k$ и $P_{2k+2}^k$ являются единственными, отличными от полных реализуемыми графами.

Условия реализуемости графов, дополнительных к деревьям, найдены в [15]. Оказалось, что класс $\overline{\mathcal{T}} = \{\overline{T} : T$ – дерево$\}$ содержит только две серии реализуемых графов, а именно $\overline{P}_n$ и $\overline{K}_{1,n}$, $n \geqslant 1$. Отметим, что в [23] построен более широкий, чем $\overline{\mathcal{T}} \setminus \{\overline{P}_n, \overline{K}_{1,n} : n \geqslant 1\}$, класс плотных графов, у которых отсутствуют реализации.

Большое число работ посвящено изучению свойств локально $\mathcal{H}$-графов в случае, когда $|\mathcal{H}| > 1$ (см. обзор в [12]). Известно, например, строение локально котриангулярных графов [20]. Граф $G$ обладает свойством *котриангулярности*, если для любой пары вершин $u$ и $v$, не смежных в $G$, существует третья вершина $w$ такая, что подграф $G(T)$, порожденный множеством $T = \{u, v, w\}$, не содержит ребер и любая вершина $x$, не лежащая в $T$, либо смежна только с одной вершиной из $T$, либо со всеми вершинами из $T$. Свойством котриангулярности обладает, например, граф, дополнительный к реберному графу полного графа $K_n$. Заметим, что $\overline{L(K_n)}$ – плотный граф.

В обзоре [21] изучались некоторые теоретико-графовые свойства реализуемых графов, содержащих доминирующие вершины. Напомним, что вершина графа, смежная с каждой другой его вершиной, называется *доминирующей.* Ясно, что если граф порядка $n$ содержит $\lceil cn \rceil$ $(0 < c \leqslant 1)$ доминирующих вершин, то он имеет не меньше $cn(cn-1)/2 + \lceil cn \rceil (n - \lceil cn \rceil)$ ребер. Данное свойство роднит графы, содержащие доминирующие вершины, с плотными графами.

Еще одно важное свойство реализуемых графов, содержащих $s$ $(s \geqslant 1)$ доминирующих вершин, состоит в том, что каждый такой граф допускает представление в виде $K_s + F[K_{s+1}]$, где $F$ – некоторый реализуемый граф без доминирующих вершин.

Основываясь на данном представлении, мы установили, что проблемы: существования, $f$-реализуемости, $i$-реализуемости остаются неразрешимыми и в том случае, если ограничиться произвольными конечными классами $\mathcal{H}$, состоящими из графов $H_1, H_2, \ldots, H_k$, у которых $|H_1| = |H_2| = \ldots = |H_k|$ и таких, что каждый $H_i$ содержит ровно $s$ доминирующих вершин.

Кроме того, установлено, что специальная проблема $i$-реализуемости неразрешима также в том случае, когда $H$ – конечный граф вида $K_s + F[K_{s+1}]$, где $F$ – произвольный связный, плоский или регулярный граф без доминирующих вершин.

Среди других интересных следствий выделим следующее: проблема единственности локально $\{K_s + F_i[K_{s+1}] : i = \overline{1,k}\}$-графа решается положительно в том и только в том случае, когда она решается положительно для локально $\{F_i : i = \overline{1,k}\}$-графа, $|F_1| = |F_2| = \ldots = |F_k|$.

Заметим, что графы, среди вершин которых есть хотя бы одна доминирующая, достаточно редки [5, с. 51]. Таким образом, результат, полученный в настоящей работе, может рассматриваться как еще один аргумент в пользу алгоритмической сложности обобщенной проблемы окружений даже для некоторых асимптотически узких классов графов.

Приведем определения основных операций над графами, которые используются в статье.

Пусть $G_i = (VG_i, EG_i)$ $(i = 1, 2)$ – два графа. *Произведением* $G_1 \times G_2$, *композицией* $G_1[G_2]$ называются графы, множеством вершин которых является декартово произведение $VG_1 \times VG_2$, а смежность вершин $(u_1, u_2)$ и $(v_1, v_2)$ задается следующими условиями:

а) для $G_1 \times G_2$ : либо $u_1 = v_1$ и $u_2v_2 \in EG_2$, либо $u_2 = v_2$ и $u_1v_1 \in EG_1$;

б) для $G_1[G_2]$ : либо $u_1v_1 \in EG_1$, либо $u_1 = v_1$ и $u_2v_2 \in EG_2$.

Пусть $G$ – граф порядка $n$, $\varphi : VG \to \{1, 2, \ldots, n\}$ – биекция, $H_1, H_2, \ldots, H_n$ – конечные графы, причем $VH_1 \cap VH_2 \cap \ldots \cap VH_n = \emptyset$. Определим операцию *расширения* графа $G$ по графам $H_1, H_2, \ldots, H_n$, обозначаемую через $G \leftarrow [H_1, H_2, \ldots, H_n]$, следующим образом:

$$V(G \leftarrow [H_1, H_2, \ldots, H_n]) = \bigcup_i VH_i$$

и две вершины $u \in H_i$, $v \in H_j$ смежны в графе $G \leftarrow [H_1, H_2, \ldots, H_n]$, если либо $i = j$ и вершины $u, v$ смежны в графе $H_i$, либо $i \neq j$ и вершины $\varphi^{-1}(i)$, $\varphi^{-1}(j)$ смежны в графе $G$.

## 1. Свойства реализуемых классов графов

Установим предварительно некоторые полезные факты.

Л е м м а 1. *Пусть* $N(G_1) = \{H_1, H_2, \ldots, H_k\}$ *и* $N(G_2) = \{F_1, F_2, \ldots, F_l\}$, *где* $G_1$ *и* $G_2$ – *конечные графы,* $k \geqslant 1$, $l \geqslant 1$. *Тогда верны следующие два равенства:*

*а)* $N(G_1[G_2]) = \{\langle F_j + H_i[G_2] \,:\, i = \overline{1,k},\, j = \overline{1,l} \rangle\}$,

*б)* $N(G_1 \times G_2) = \{\langle H_i \cup F_j \,:\, i = \overline{1,k},\, j = \overline{1,l} \rangle\}$.

*В частности, если* $G_2$ – *реализация графа* $F$, *то* $N(G_1[G_2]) = \{F + H_i[G_2] \,:\, i = \overline{1,k}\}$, $N(G_1 \times G_2) = \{F \cup H_i \,:\, i = \overline{1,k}\}$.

Доказательство. Зададим на $VG_\alpha$ бинарное отношение $\sigma_\alpha$ $(\alpha = 1, 2)$ : будем писать $u\,\sigma_\alpha\,v$ если и только если $G_\alpha(N(u)) \cong G_\alpha(N(v))$. Очевидно, $\sigma_\alpha$ – отношение эквивалентности. Следовательно, $\sigma_\alpha$ разбивает $VG_\alpha$ на классы эквивалентных элементов. Положим

$$VG_1/\sigma_1 = \{U_i \,:\, i = \overline{1,k}\}, \quad VG_2/\sigma_2 = \{V_j \,:\, j = \overline{1,l}\},$$

где $G_1(N(u)) \cong H_i$ для каждой вершины $u \in U_i$ и $G_2(N(v)) \cong F_j$ для каждой вершины $v \in V_j$. С помощью $\sigma_1$ и $\sigma_2$ построим эквивалентность $\sigma_3$ на $VG_1[G_2]$, положив $(u_1, u_2)\,\sigma_3\,(v_1, v_2)$, если $u_1\,\sigma_1\,v_1$ и $u_2\,\sigma_2\,v_2$. Очевидно, $VG_1[G_2]/\sigma_3 = \{U_i \times V_j \,:\, i = \overline{1,k},\, j = \overline{1,l}\}$, а так как для $(u, u') \in VG_1[G_2]$

$$N((u, u'), G_1[G_2]) = (N(u, G_1) \times VG_2) \cup (\{u\} \times N(u', G_2)),$$

то отображение $VG_1[G_2]/\sigma_3 \to N(G_1[G_2])$, при котором $U_i \times V_j \mapsto F_j + H_i[G_2]$ есть сюръекция. Отсюда и вытекает справедливость равенства а).

Повторяя предыдущие рассуждения и учитывая, что

$$N((u, u'), G_1 \times G_2) = (N(u, G_1) \times \{u'\}) \cup (\{u\} \times N(u', G_2)),$$

приходим к равенству б).

Для доказательства равенств а), б) в случае $N(G_2) = \{F\}$ достаточно лишь заметить, что графы $F \cup H_i$ и $F \cup H_j$, а также $F + H_i[G_2]$ и $F + H_j[G_2]$ изоморфны тогда и только тогда, когда изоморфны графы $H_i$ и $H_j$. Лемма 1 доказана.

Заметим, что в общем случае реализуемость дизъюнктного объединения $H \cup F$ не является достаточной для существования графов $G_1$ и $G_2$ таких, что $N(G_1) = \{H\}$, $N(G_2) = \{F\}$. Примером такого графа с наименьшим числом вершин служит $P_3 \cup K_2$. Действительно, отсутствие каких-либо реализаций для графа $P_3$ вытекает, например, из следствия 1, а один из возможных способов построения целого семейства локально $P_3 \cup K_2$-графов представлен на рис. 2. Этот способ основан на использовании срединно геометрического графа $M(L)$ любого плоского 4-регулярного 2-связного графа $L$, не содержащего граней, ограниченных меньше, чем четырьмя ребрами (определение графа $M(L)$ см. в [25, с. 124]).

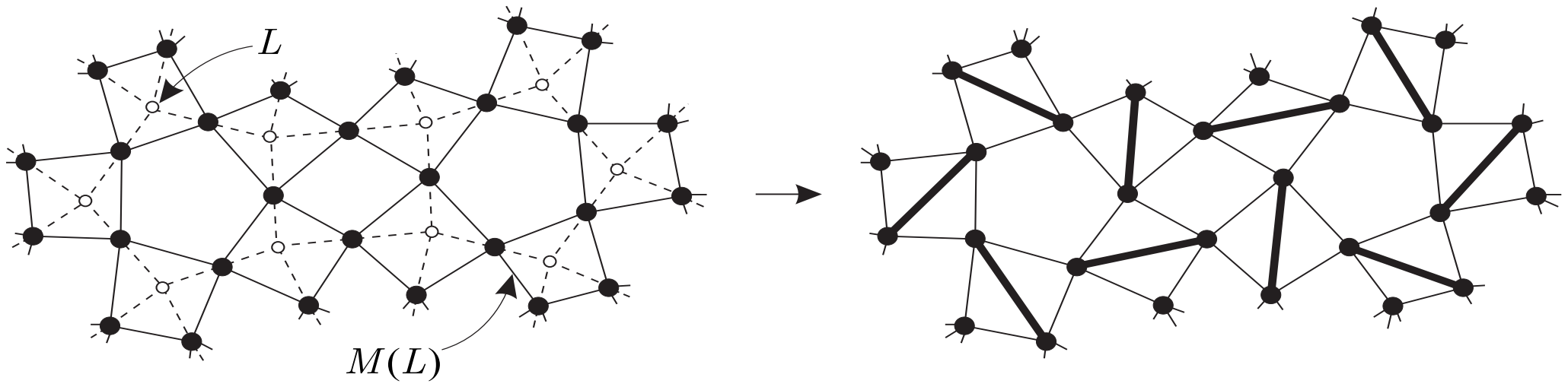

Рис. 2. Фрагмент построения реализации графа $P_3 \cup K_2$.

Рассмотренный пример по существу служит первым проявлением более общего результата, согласно которому, каждый конечный граф является подграфом некоторого конечного реализуемого графа [2, 21].

Ниже окажется полезным следующий результат.

Лемма 2. *Для любых натуральных чисел $r$ и $n$, таких, что $1 \leqslant r \leqslant n$, существует двудольный $r$-регулярный граф порядка $2n$.*

Доказательство. Будем доказывать индукцией по числу $n$. При $n = 1, 2$ лемма тривиальна. Пусть $n > 2$ и для $n - 1$ и всех $r$, таких, что $1 \leqslant r \leqslant n - 1$, пусть лемма верна. Покажем, что она верна для числа $n$ и для всех чисел $r$, подчиненных неравенствам $1 \leqslant r \leqslant n$.

Пусть $G = (X, Y, E)$ – двудольный $r$-регулярный граф порядка $2(n - 1)$, где $r$ – произвольное число, $1 \leqslant r \leqslant n - 1$. Выберем в этом графе $r$ попарно несмежных ребер $e_1, e_2, \ldots, e_r$, где $e_i = u_i v_i$, $u_i \in X$, $v_i \in Y$ $(i = \overline{1,r})$. Это всегда можно сделать, так как в любом непустом регулярном двудольном графе существует совершенное паросочетание (множество попарно несмежных ребер графа, покрывающих все его вершины). Рассмотрим граф $G^*$, у которого $VG^* = X \cup Y \cup \{u, v\}$. В множество ребер этого графа включим, во-первых, все ребра графа $G$, отличные от $e_1, e_2, \ldots, e_r$, а во-вторых, все ребра вида $uv_i$ и вида $vu_i$ $(i = \overline{1, r})$. Легко видеть, что построенный таким образом граф $G^*$ имеет порядок $2n$ и обладает нужными свойствами. Если $r = n$, то мы вправе считать, что $G^* \cong K_{n,n}$. Лемма 2 доказана.

Следующая лемма указывает на тесную связь, существующую между реализуемостью классов $\{H_1, H_2, \ldots, H_k\}$ и $\{\langle H_1 \cup O_{n_1}, H_2 \cup O_{n_2}, \ldots, H_k \cup O_{n_k}\rangle\}$, где $n_1, n_2, \ldots, n_k$ – произвольные натуральные числа, подчиненные условию

$$|H_1| + n_1 = |H_2| + n_2 = \ldots = |H_k| + n_k. \tag{1}$$

Лемма 3. *Для конечных попарно неизоморфных графов $H_1, H_2, \ldots, H_k$ $(k \geqslant 1)$ следующие два утверждения равносильны:*

*i) существует конечный граф $G$, у которого $N(G) = \{H_1, H_2, \ldots, H_k\}$;*

*ii) для любых натуральных чисел $n_1, n_2, \ldots, n_k$, подчиненных условию (1), существует такой конечный регулярный граф $G_0$, что $N(G_0) = \{\langle H_1 \cup O_{n_1}, H_2 \cup O_{n_2}, \ldots, H_k \cup O_{n_k}\rangle\}$.*

Доказательство. $i) \Rightarrow ii)$. Пусть $p$ – максимальная, а $q$ – минимальная из степеней вершин графа $G$, $n = p - q$. Пусть, далее, $G^* \cong G \times K_{n+m,n+m}$, где $m$ – произвольное натуральное число. Тогда так как окружение каждой вершины в $K_{n+m,n+m}$ порождает $O_{n+m}$, то, согласно лемме 1, $N(G^*) = \{H_i \cup O_{n+m} : i = \overline{1, k}\}$. Мы построим преобразование $\pi$ графа $G^*$ в регулярный граф $G_0$ такое, что $N(G_0) = \{\langle H_1 \cup O_{n_1}, \ldots, H_k \cup O_{n_k}\rangle\}$.

Для этого определим на $VG^*$ бинарное отношение $\sim$, положив $(u_1, u_2) \sim (v_1, v_2)$ для вершин $(u_1, u_2)$ и $(v_1, v_2)$, если $u_1 = v_1$. Очевидно, что это отношение есть эквивалентность. Следовательно, мы получили разбиение множества $VG^*$ на классы, отнеся в один класс все вершины, эквивалентные друг другу. Пусть

$$VG^* = V_1 \cup V_2 \cup \ldots \cup V_t \tag{2}$$

– такое разбиение. Легко проверяются следующие свойства классов $V_j$, $j = \overline{1, t}$ : число $t$ этих классов равно порядку графа $G$, $G^*(V_j) \cong K_{n+m,n+m}$ и, наконец, для любых двух вершин $(u_1, u_2), (v_1, v_2) \in V_j$ имеет место изоморфизм $G(N(u_1)) \cong G(N(v_1))$.

Пусть $V_j$ – произвольный класс разбиения (2) и $(u, u')$ – произвольная вершина из этого класса. Положим

$$\delta_j = \deg_G u - q.$$

Тогда на основании леммы 2 $G^*(V_j)$ содержит $\delta_j$-регулярный подграф $F_j$ порядка $2(m+n)$. В частности, при $\delta_j = 0$ считаем $VF_j = VG^*(V_j)$, $EF_j = \emptyset$. Выполним теперь следующее преобразование графа $G^*$ : удалим в этом графе те и только те ребра подграфа $G^*(V_j)$, которые вошли в множество $EF_j$. Легко видеть, что в полученном в результате такого преобразования графе, окружение вершины $(u, u')$ будет порождать $G_u \cup O_{n_j}$, где $G_u \cong G(N(u))$ и $n_j = n + m - \delta_j$. В частности, отсюда и из предыдущего выводим

$$|G_u| + n_j = |G_u| + n + m - \delta_j = |G_u| + n + m - (\deg_G u - q) = n + m + q = m + p. \tag{3}$$

Применив последовательно указанное преобразование к каждому из оставшихся классов разбиения (2), получим, согласно (3), конечный $(m+p)$-регулярный граф, который обозначим через $\pi(G^*)$. Из способа построения графа $\pi(G^*)$ ясно, что

$$N(\pi(G^*)) = \{\langle G_u \cup O_{n_j} : (u, u') \in V_j,\ j = \overline{1, t}\rangle\} = \{\langle H_i \cup O_{n_i} : i = \overline{1, k}\rangle\},$$

где $|H_i| + n_i = m + p$, $i = \overline{1, k}$. Следовательно, полагая $G_0 \cong \pi(G^*)$ и учитывая произвольность параметра $m$, получим требуемое.

*ii)* $\Rightarrow$ *i)*. Обратно, пусть для любых натуральных чисел $n_1, \ldots, n_k$, подчиненных условию (1), существует такой конечный граф $G_0$, что $N(G_0) = \{\langle H_1 \cup O_{n_1}, \ldots, H_k \cup O_{n_k} \rangle\}$. Возможны два случая: а) графы $H_1, \ldots, H_k$ не имеют изолированных вершин, б) по крайней мере один $H_i$ $(1 \leqslant i \leqslant k)$ содержит изолированные вершины.

Рассмотрим сперва случай а). Наша цель – построить из графа $G_0$ такой граф $G$, что $N(G) = \{H_1, \ldots, H_k\}$. Заметим сразу, что графы $H_1 \cup O_{n_1}, \ldots, H_k \cup O_{n_k}$ попарно неизоморфны. Поэтому если через $H'_i$ обозначить подграф графа $H_i \cup O_{n_i}$, порожденный множеством всех его неизолированных вершин, то окажется, что $H'_i \cong H_i$, $i = \overline{1,k}$. Удалив теперь из $G_0$ все *особые* ребра, т.е. ребра, не входящие ни в один треугольник (цикл $C_3$), мы получим граф $G'$, который, очевидно, является конечной реализацией класса $\{H'_i : i = \overline{1,k}\}$. Последнее в силу изоморфизма $H'_i \cong H_i$ $(i = \overline{1,k})$ дает нам право считать, что $G \cong G'$.

Перейдем теперь к случаю б). В этом случае мы не будем проводить точную, но громоздкую процедуру построения графа $G$. Существование такого графа легко усмотреть из следующего. По списку $H_1 \cup O_{n_1}, \ldots, H_k \cup O_{n_k}$, среди графов которого могут быть изоморфные, сформируем список $\mathcal{H}'$, состоящий из графов $H'_1, \ldots, H'_k$. Здесь $H'_i$, как и раньше, – подграф, порожденный множеством всех неизолированных вершин графа $H_i \cup O_{n_i}$, $i = \overline{1,k}$. Не исключено, что при некотором $i$ $H'_i$ – граф с пустым множеством вершин. В таком случае по определению положим $H'_i \cong K_0$, причем считаем, что $K_1$ – связная реализация графа $K_0$.

Не уменьшая общности, предположим, что первые $l$ $(l \leqslant k)$ графов, входящие в список $\mathcal{H}'$, попарно неизоморфны. Очевидно, удалив из $G_0$ все особые ребра, получим граф $G'$, у которого $N(G') = \{H'_i : i = \overline{1,l}\}$. Без потери общности рассуждений можно считать, что $G'$ – нетривиальный граф. Свяжем теперь с каждым графом $H'_i$ $(i = \overline{1,l})$ целое неотрицательное $n'_i$ – число графов из $\mathcal{H}'$, которые изоморфны $H'_i$. Пусть

$$n = \max\{n'_1, \ldots, n'_l\}, \quad m = \max\{m_i : H_i \cong H'_i \cup O_{m_i}, i = \overline{1,k}\}.$$

Рассмотрим граф $G^*$, состоящий из $n$ копий декартова произведения $G' \times K_{m,m}$. Ясно, что $G^*$ – конечная реализация класса $\{H'_i \cup O_m : i = \overline{1,l}\}$. В тоже время в графе $G^*$ с точностью до удаления некоторого подмножества особых ребер можно выделить такой остовный подграф $G$, что $N(G) = \{H_1, \ldots, H_k\}$. Очевидно, это возможно. Лемма 3 доказана.

Пусть $\mathcal{F} = \{F_1, F_2, \ldots, F_k\}$ – произвольный конечный класс графов без доминирующих вершин и таких, что $|F_1| = |F_2| = \ldots = |F_k|$. Тогда верна

Л е м м а 4. *Невозможен алгоритм, распознающий по классу $\mathcal{F}$, существует ли конечный регулярный граф $G$, реализующий $\mathcal{F}$.*

Д о к а з а т е л ь с т в о. Пусть, напротив, возможен алгоритм $\mathcal{A}$, определяющий по произвольному классу $\mathcal{F}$, подчиненному условиям леммы, существует ли конечный регулярный локально $\mathcal{F}$-граф $G$. Рассмотрим произвольный конечный класс $\mathcal{H}$, состоящий из графов $H_1, \ldots, H_k$. На основании леммы 3, мы вправе считать, что конечный граф $G^*$, реализующий $\mathcal{H}$, существует тогда и только тогда, когда для любых натуральных чисел $n_1, \ldots, n_k$, подчиненных условию (1), существует такой конечный регулярный граф $G_0$, что

$$N(G_0) = \{\langle H_1 \cup O_{n_1}, \ldots, H_k \cup O_{n_k} \rangle\}.$$

Следовательно, подавая на вход алгоритма $\mathcal{A}$ класс $\{\langle H_1 \cup O_{n_1}, \ldots, H_k \cup O_{n_k} \rangle\}$, мы можем установить существование графа $G_0$, а значит, и существование графа $G^*$. Но последнее невозможно, ибо, согласно теореме 2, проблема $f$-реализуемости алгоритмически неразрешима. Лемма 4 доказана.

Отметим еще одну любопытную связь между реализуемостью классов $\{H_1, H_2, \ldots, H_k\}$ и $\{n_1 H_1 \cup n_2 H_2 \cup \ldots \cup n_k H_k\}$, где $n_1, n_2, \ldots, n_k$ – произвольные натуральные числа, а $nH$ – граф с $n$ компонентами, каждая из которых изоморфна $H$.

Л е м м а 5. *Для конечных попарно неизоморфных связных графов $H_1, H_2, \ldots, H_k$ $(k \geqslant 1)$ следующие два утверждения равносильны:*

*i) существует граф $G$, имеющий не более чем счетное множество вершин, у которого $N(G) = \{H_1, H_2, \ldots, H_k\}$;*

*ii) для любых натуральных чисел* $n_1, n_2, \dots, n_k$ *существует такой связный бесконечный регулярный граф* $G_0$, *что* $N(G_0) = \{H\}$, *где* $H \cong n_1H_1 \cup n_2H_2 \cup \ldots \cup n_kH_k$.

Доказательство. $i) \Rightarrow ii)$. Рассмотрим подробнее случай, когда $G$ – связный конечный граф, так как процедура построения графа $G_0$ в этом случае упрощается.

Пусть $G$ – такой граф и $H \cong n_1H_1 \cup \ldots \cup n_kH_k$, где $n_1, \dots, n_k$ – произвольные натуральные числа. Для доказательства существования связного бесконечного графа $G_0$, у которого $N(G_0) = \{H\}$, сначала индуктивно построим последовательность $S = (G_1, G_2, \dots, G_i, \dots)$ связных конечных графов $G_i$ таких, что $N(G_1) = N(G)$ и $N(G_i) = N(G) \cup \{H\}$ для $i \geqslant 2$. Положим $G_1 \cong G$. Пусть граф $G_i$ уже построен, $i \geqslant 1$, и пусть все вершины его занумерованы последовательными натуральными числами, а окружение каждой $u \in VG_i$ порождает подграф, изоморфный либо графу $H$, либо графу $H_j$ для некоторого $j \in \{1, \dots, k\}$, причем существует такая вершина $v \in VG_i$, что $G_i(N(v))$ не изоморфен $H$.

Выберем в графе $G_i$ вершину $w$ с наименьшим номером и такую, что $G_i(N(w))$ не изоморфен $H$. Не исключая общности рассуждений, будем считать, что $G_i(N(w)) \cong H_k$. К вершине $w$ "приклеим"

$$n = n_1 + n_2 + \ldots + n_k - 1$$

клик $Q_1^i, \dots, Q_n^i$ порядка $|G|$, для каждой из которых $w$ – единственная вершина, общая с графом $G_i$ (см. рис. 3а). Любые две из этих клик не должны иметь общих ребер. Рассмотрим теперь граф $Q$, состоящий из вершин и ребер клик $Q_1^i, \dots, Q_n^i$, и занумеруем элементы множества $VQ \setminus \{w\}$ последовательными натуральными числами, превосходящими максимальный среди номеров вершин графа $G_i$.

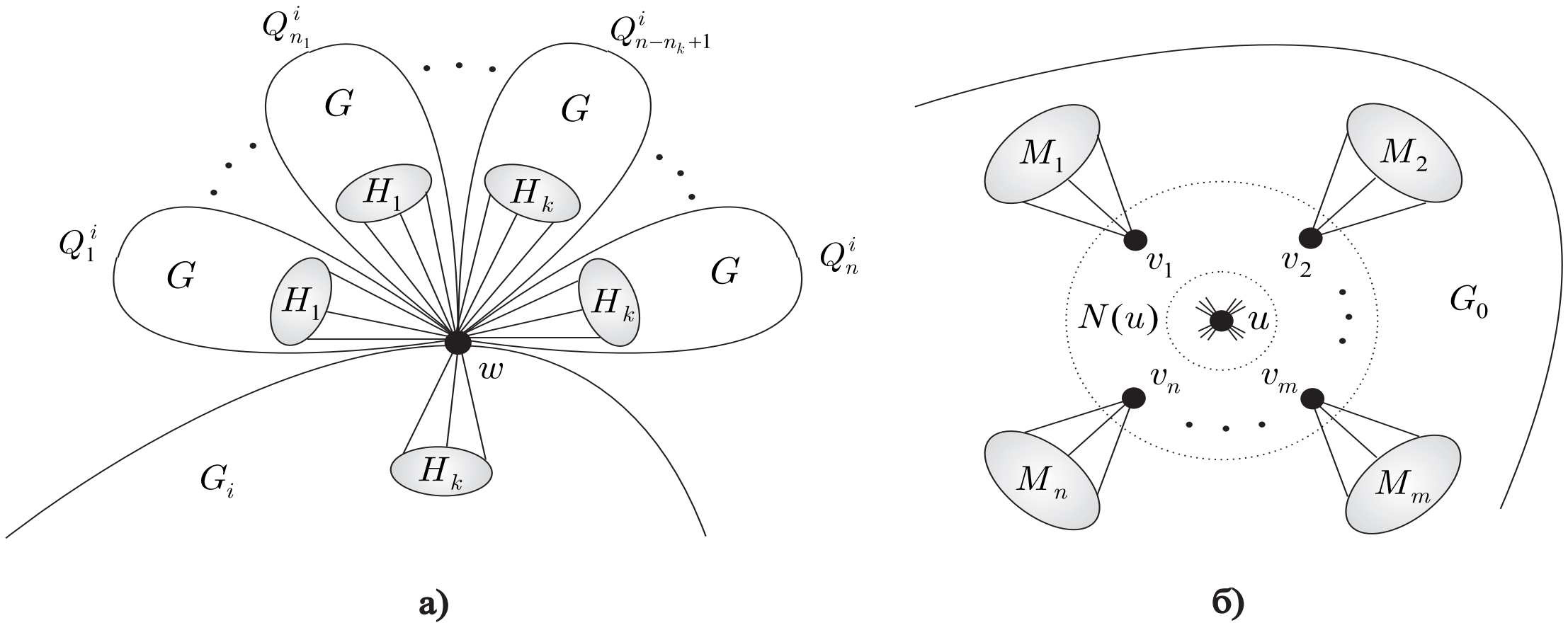

Рис. 3. Рисунок к доказательству леммы 7.

Пусть $Q_m^i$ $(1 \leqslant m \leqslant n)$ – произвольная максимальная клика из $Q$. Выберем в этой клике остовный подграф $F_m^i$ такой, что $F_m^i \cong G$ и $F_m^i(N(w)) \cong H_j$, где $j = 1$ при $1 \leqslant m \leqslant n_1$, а во всех остальных случаях индекс $j$ подчинен неравенствам

$$n_1 + n_2 + \ldots + n_{j-1} < m \leqslant n_1 + n_2 + \ldots + n_j. \tag{4}$$

Выполним следующее преобразование графа $Q$ : удалим в этом графе те и только те ребра подграфа $Q(VQ_m^i)$, которые не вошли во множество $EF_m^i$. Применив последовательно указанное преобразование к каждой из оставшихся максимальных клик графа $Q$, мы получим связный конечный граф, который обозначим через $G_{i+1}$, причем

$$VG_{i+1} = VG_i \cup VQ_1^i \cup \ldots \cup VQ_n^i, \quad EG_{i+1} = EG_i \cup EF_1^i \cup \ldots \cup EF_n^i.$$

Как легко видеть, $G_{i+1}(N(w)) \cong H$. В тоже время для каждой вершины $u \in VG_{i+1}$, $u \neq w$, $G_{i+1}(N(u)) \cong G_i(N(u))$, и если $v \in (VQ_1^i \cup \ldots \cup VQ_n^i) \setminus \{w\}$, то $G_{i+1}(N(v)) \cong H_j$ для некоторого $j \in \{1, \dots, k\}$. Из последнего наблюдения и выполненных построений вытекает, что искомым графом $G_0$ является граф со счетным множеством вершин $VG_0$ и счетным множеством ребер $EG_0$ :

$$VG_0 = \bigcup_{i=1}^{\infty} VG_i, \quad EG_0 = \bigcup_{i=1}^{\infty} EG_i.$$

Действительно, для каждой $u \in VG_0$ можно указать такое наименьшее число $i_0$, что $u \in VG_i$ для всех $i \geqslant i_0$. Следовательно, вершина $u$ имеет некоторый номер

$$l \in \big\{|G| + n(|G|-1)(i_0-1) - \varepsilon \,:\, \varepsilon = 0, 1, \ldots, n(|G|-1)-1\big\}.$$

Но тогда по построению $G_{l+1}(N(u)) \cong H$, а так как в дальнейшем окружение вершины $u$ сохраняется, то $G_0$ – искомый граф.

*ii*) $\Rightarrow$ *i*). Пусть $G_0$ – такой бесконечный регулярный граф, что $N(G_0) = \{n_1H_1 \cup \ldots \cup n_kH_k\}$, где $n_1, \ldots, n_k$ – натуральные числа. Для доказательства существования графа $G$ поступим следующим образом. Возьмем произвольную вершину $u \in VG_0$ и разобьем ее окружение на

$$n = n_1 + n_2 + \ldots + n_k$$

частей $M_1, \ldots, M_n$ так, чтобы $G_0(M_m) \cong H_j$, где, как и выше, $j = 1$ при $1 \leqslant m \leqslant n_1$, а во всех остальных случаях индекс $j$ подчинен неравенствам (4). Выполним теперь следующее преобразование графа $G_0$ : удалим вершину $u$ вместе с инцидентными ей ребрами, добавим новые вершины $v_1, \ldots, v_n$, вершину $v_1$ соединим ребром с каждой вершиной из множества $M_1$, вершину $v_2$ – с каждой вершиной из множества $M_2$ и т. д. (см. рис. 3б). Легко видеть, что в полученном в результате такого преобразования графе, окружение вершины $v_m$ будет порождать подграф, изоморфный $H_j$ (здесь индексы $m$ и $j$ те же, что и раньше), а окружение каждой $v \in VG_0 \backslash \{u\}$ сохраняется, в том смысле, что подграф, порожденный этим окружением изоморфен $H$. Применив последовательно указанное преобразование к каждой из оставшихся вершин графа $G_0$, мы, очевидно, получим граф $G$, у которого $N(G) = \{H_1, \ldots, H_k\}$. Лемма 5 доказана.

## 2. Реализуемые классы графов с доминирующими вершинами

В этом разделе будет доказано, что задачи 1 – 4 для произвольных конечных классов графов сводятся к тем же задачам для классов, состоящих из графов $H_1, H_2, \ldots, H_k$, у которых $|H_1| = |H_2| = \cdots = |H_k|$ и таких, что каждый $H_i$ содержит ровно $s$ доминирующих вершин. Структурное свойство этих классов, на котором основано сведение, описывает

Теорема 3. *Пусть $H_1, H_2, \ldots, H_k$ $(k \geqslant 1)$ – конечные попарно неизоморфные графы, $|H_1| = |H_2| = \cdots = |H_k|$. Пусть, далее, каждый граф $H_i$ содержит ровно $s$ доминирующих вершин $(s \geqslant 1)$. Тогда следующие два утверждения эквивалентны:*

*i) существует конечный граф $G$, у которого $N(G) = \{H_1, H_2, \ldots, H_k\}$;*

*ii) существуют такой конечный граф $G_0$ и такие конечные попарно неизоморфные графы без доминирующих вершин $F_1, F_2, \ldots, F_k$, $|F_1| = |F_2| = \ldots = |F_k|$, что $G \cong G_0[K_{s+1}]$ и $N(G_0) = \{F_1, F_2, \ldots, F_k\}$. При этом $H_i \cong K_s + F_i[K_{s+1}]$ для каждого $i = 1, 2, \ldots, k$.*

Доказательство. *i*) $\Rightarrow$ *ii*). Пусть $G$ и $H_1, \ldots, H_k$ – произвольные графы, удовлетворяющие условиям теоремы. Зададим на множестве $VG$ бинарное отношение $\sigma$ : будем писать $u\,\sigma\,v$ тогда и только тогда, когда

$$N(u) \cup \{u\} = N(v) \cup \{v\}.$$

Очевидно, $\sigma$ – отношение эквивалентности. Пусть, далее, $VG/\sigma = \{V_1, \ldots, V_p\}$ – фактор-множество множества $VG$ по эквивалентности $\sigma$. Просто проверяются следующие свойства смежных классов из $VG/\sigma$ :

а) $G(V_i)$ – полный граф, $i = \overline{1,p}$;

б) если вершина $u \in V_i$ смежна с вершиной $v \in V_j$, то $u$ смежна со всеми вершинами множества $V_j$.

Пусть теперь граф $G_\sigma$, разбиение $VG/\sigma$ и биекция $\varphi : VG_\sigma \to VG/\sigma$ таковы, что $uv \in$ $\in EG_\sigma$ если и только, если $G(\varphi(u) \cup \varphi(v))$ – клика, т.е. $G_\sigma$ – граф соседства классов из $VG/\sigma$. Тогда для $G$ в силу свойств а), б) имеем

$$G \cong G_\sigma \leftarrow [K_{n_1}, K_{n_2}, \ldots , K_{n_p}],$$

где $n_i = |V_i|$, $i = \overline{1,p}$.

Так как $G$ – регулярный граф, то эквивалентность $\sigma$ на $VG$ можно определить и так: $u\,\sigma\,v$ тогда и только тогда, когда вершина $u$ доминирует в окрестности вершины $v$, т.е. в $G(N(v) \cup \{v\})$. Таким образом, каждое множество $V_i$ можно рассматривать как класс эквивалентных вершин, доминирующих в окрестности некоторой вершины из этого же класса. Последнее означает, что $V_i = \{u_i\} \cup D(G(N(u_i)))$, где $u_i$ – произвольная вершина из $V_i$, а $D(H)$ – множество доминирующих вершин графа $H$. Отсюда и из условий теоремы вытекают равенства $|V_i| = s + 1$, $i = \overline{1,p}$. Следовательно,

$$G \cong G_\sigma \leftarrow [K_{s+1}, K_{s+1}, \ldots , K_{s+1}] \cong G_\sigma[K_{s+1}].$$

Полагая теперь $G_0 \cong G_\sigma$, получим $G \cong G_0[K_{s+1}]$.

По условию $N(G) = \{H_1, \ldots , H_k\}$. Следовательно, в $G$ есть такие вершины $u_1, \ldots , u_k$, что $G(N(u_i)) \cong H_i$, $i = \overline{1,k}$. Рассмотрим отображение

$$\psi : N(G) \to N(G_0), \quad H_i \mapsto F_i \cong G_0(N((\varphi^{-1}\pi)(u_i))), \tag{5}$$

где $\pi : VG \to VG/\sigma$ – каноническое отображение $VG$ на $VG/\sigma$, т.е. $\pi$ каждой вершине графа $G$ ставит в соответствие класс эквивалентности из $VG/\sigma$, в котором эта вершина содержится. Покажем, что $\psi$ – биекция. Действительно, в силу предыдущего и с учетом леммы 1 для каждого графа $H \cong G(N(u))$, $u \in VG$ справедливо представление

$$H \cong K_s + F[K_{s+1}], \tag{6}$$

где $F \cong G_0(N((\varphi^{-1}\pi)(u)))$. Отсюда легко заключить, что $\psi$ сюръективно. С другой стороны, если графы $H_i$, $H_j$ не изоморфны, то в силу (6) не изоморфными будут также и графы $F_i[K_{s+1}]$, $F_j[K_{s+1}]$. Следовательно, $F_i$ не изоморфен $F_j$ при $i \neq j$, т.е. $\psi$ – инъекция.

Итак, $N(G_0) = \{F_1, \ldots , F_k\}$ и $H_i \cong K_s + F_i[K_{s+1}]$ $(i = \overline{1,k})$, где соответствие между графами $H_i$ и $F_i$ определено отображением (5). Наконец, из (6) вытекает, что $F_i$ не имеет доминирующих вершин, ибо в противном случае граф $H_i \cong K_s + F_i[K_{s+1}]$ содержал бы более $s$ таких вершин.

*ii*) ⇒ *i*). Достаточно воспользоваться леммой 1 и представлением $G \cong G_0[K_{s+1}]$. Теорема 3 доказана.

Частным случаем общей теоремы 3 является

Следствие 1. *Пусть выполнены условия теоремы 3. Тогда получаем необходимые условия существования графа $G$ в следующем виде:* $\deg_{H_i} u \geqslant 2s$ *и* $\deg_{H_i} u \equiv 2s(\mathrm{mod}(s+1))$ *для каждой вершины* $u \in VH_i \setminus D(H_i)$ *и каждого* $i = 1, 2, \ldots , k$.

В связи с доказательством теоремы 3 отметим еще одно обстоятельство. Если для конечных графов $F_1, F_2, \ldots , F_k$, удовлетворяющих условию *ii*) теоремы, явно описан класс

$$\mathcal{F} = \{G_0 : N(G_0) = \{F_1, F_2, \ldots , F_k\}\},$$

то явное описание допускает и класс

$$\mathcal{G} = \{G : N(G) = \{H_1, H_2, \ldots , H_k\}\},$$

где $H_i \cong K_s + F_i[K_{s+1}]$ $(i = \overline{1,k})$, $s$ – произвольное натуральное число. Действительно, как легко видеть, изоморфизм $G \cong G_0[K_{s+1}]$ задает биекцию $\mathcal{G} \to \mathcal{F}$, $G \mapsto G_0[K_{s+1}]$. В частности, отсюда следует, что проблемы единственности конечных локально $\{H_1, H_2, \ldots, H_k\}$-графа и локально $\{F_1, F_2, \ldots, F_k\}$-графа эквивалентны.

Следующая теорема фактически утверждает, что проблема существования, т.е. когда нет явных ограничений на конечность или бесконечность реализаций, неразрешима.

Т е о р е м а 4. *Невозможен алгоритм, определяющий по произвольному конечному классу связных графов $\mathcal{H}$, существует ли локально $\mathcal{H}$-граф, имеющий не более, чем счетное множество вершин.*

Д о к а з а т е л ь с т в о. Пусть вопреки теореме такой алгоритм $\mathcal{B}$ существует. Рассмотрим произвольный конечный класс $\mathcal{H}$, состоящий из связных графов $H_1, \ldots, H_k$. В силу леммы 5 мы можем считать, что локально $\mathcal{H}$-граф, имеющий не более, чем счетное множество вершин, существует тогда и только тогда, когда для любых натуральных чисел $n_1, \ldots, n_k$ существует связная бесконечная реализация графа

$$H \cong n_1 H_1 \cup \ldots \cup n_k H_k. \tag{7}$$

Таким образом, подавая на вход алгоритма $\mathcal{B}$ класс $\mathcal{H}$, мы установим существование локально $\mathcal{H}$-графа и, следовательно, существование связного локально $H$-графа. Но последнее противоречит неразрешимости специальной проблемы $i$-реализуемости в классе всех несвязных графов (см. [3]), ибо представление (7) имеет место для любого конечного несвязного графа $H$. Теорема 4 доказана.

В следующей теореме графы $H_1, H_2, \ldots, H_k$ те же, что в теореме 3.

Т е о р е м а 5. *Для произвольного конечного класса графов $\mathcal{H} = \{H_1, H_2, \ldots, H_k\}$ проблемы: существования, $f$-реализуемости, $i$-реализуемости алгоритмически неразрешимы.*

Д о к а з а т е л ь с т в о. Рассмотрим сперва случай $f$-реализуемости. Предположим от противного, что возможен алгоритм $\mathcal{C}$, корректно распознающий по произвольному конечному классу $\mathcal{H} = \{H_1, \ldots, H_k\}$, подчиненному условиям теоремы 3, существует ли конечный локально $\mathcal{H}$-граф. Пусть $F_1, \ldots, F_k$ – произвольные попарно неизоморфные графы без доминирующих вершин и такие, что $|F_1| = \ldots = |F_k|$. Построим графы

$$K_s + F_1[K_{s+1}], \ldots, K_s + F_k[K_{s+1}].$$

Ясно, что порядки этих графов равны и каждый из них содержит ровно $s$ доминирующих вершин. Таким образом, подавая на вход алгоритма $\mathcal{C}$ класс $\mathcal{K} = \{K_s + F_i[K_{s+1}] : i = \overline{1,k}\}$, мы установим существование конечного локально $\mathcal{K}$-графа и, следовательно, согласно теореме 3, конечного локально $\mathcal{F}$-графа, где $\mathcal{F} = \{F_1, \ldots, F_k\}$. Итак, алгоритм $\mathcal{C}$ корректно распознает конечные локально $\mathcal{F}$-графы. Получили противоречие с леммой 4.

Перейдем к случаю $i$-реализуемости. Пусть все графы класса $\mathcal{H}$ изоморфны $H$. Точно так же, как и в лемме 4, устанавливается, что специальная проблема $i$-реализуемости неразрешима для произвольного конечного графа $F$ без доминирующих вершин. Следовательно, по аналогии со случаем $f$-реализуемости, невозможен алгоритм, устанавливающий по $\mathcal{H}$ существование бесконечного локально $H$-графа для $H \cong K_s + F[K_{s+1}]$.

Эту аналогию легко проследить, если заметить, что теорема 3 остается верной и в том случае, когда $G$ – бесконечная связная реализация класса $\mathcal{H} = \{H\}$. В частности, из проведенных рассуждений следует, что специальная проблема $i$-реализуемости неразрешима также в том случае, когда $H$ – конечный граф вида $K_s + F[K_{s+1}]$, где $F$ – произвольный связный, плоский или регулярный граф без доминирующих вершин (см. [3]).

Рассмотрим теперь проблему существования. Как нетрудно проследить, утверждения леммы 4 и теоремы 3 верны также в случае, когда $G$ – граф с не более, чем счетным множеством вершин. Поэтому, повторив рассуждения из случая $f$-реализуемости и учитывая теорему 4, получим утверждение о том, что проблема существования для класса $\mathcal{H}$ алгоритмически неразрешима. Теорема 5 доказана.

Пусть $\lambda$ – целая константа, $\lambda > 1$. Класс $\mathcal{D}$ графов назовем $\lambda$*-близким* к плотному, а сами графы $G \in \mathcal{D}$ – $\lambda$*-близкими* к плотным, если выполняются следующие условия:

а) существует такая функция $\mathcal{E} : \mathbb{N} \to \mathbb{N}$, что

$$\lim_{n\to\infty} \frac{n^{2-1/\lambda}}{\mathcal{E}(n)} = c,$$

где $c$ – константа, $c \geqslant 0$;

б) число ребер каждого графа $G \in \mathcal{D}$ порядка $n$ не меньше $\mathcal{E}(n)$.

Непосредственно из теоремы 5 вытекает

Следствие 2. *Для произвольного конечного класса графов $\lambda$-близких к плотным проблемы: существования, $f$-реализуемости и $i$-реализуемости алгоритмически неразрешимы.*

Доказательство. Пусть $N$ – произвольное, а $\mu$ – фиксированное натуральные числа. Пусть, далее, $F_1, \ldots, F_k$ – произвольные попарно неизоморфные графы без доминирующих вершин, причем $|F_1| = \ldots = |F_k| = N$. Выберем натуральное $n$ так, что $N = n^{1/\mu}$ и построим графы $H_i \cong K_n + F_i[K_{n+1}]$, $i = \overline{1,k}$. Для порядка графа $H_i$, очевидно, верно асимптотическое равенство

$$|H_i| \sim n^{1+1/\mu}, \quad n \to \infty.$$

Ясно также, что каждый граф $H_i$ содержит ровно $n$ доминирующих вершин. Следовательно, согласно теореме 5, для класса $\{H_i : i = \overline{1,k}\}$ проблемы существования, $f$-реализуемости и $i$-реализуемости алгоритмически неразрешимы.

Возьмем теперь натуральное число $\lambda = \mu + 1$ и покажем, что класс $\{H_i : i = \overline{1,k}\}$ является $\lambda$-близким к плотному. Как легко проверить,

$$2|EH_i| = 3n(n+1)n^{1/\mu} + 2(n+1)^2|EF_i| + n(n-1).$$

Используя последнее равенство, положим

$$\mathcal{E}(n) = \frac{3}{2}n(n+1)n^{1/\mu} + (n+1)^2\mathcal{M} + \frac{1}{2}n(n-1), \tag{8}$$

где $\mathcal{M} = \min\limits_{i=\overline{1,k}} |EF_i|$. Не ограничивая общности рассуждений, можно считать, что

$$\mathcal{M} \sim \gamma n^{1/\mu}, \quad n \to \infty, \tag{9}$$

где $\gamma$ – постоянная. Из (8) и (9) получаем асимптотику для функции $\mathcal{E}(n)$

$$\mathcal{E}(n) \sim \left(\frac{3}{2} + \gamma\right) n^{2+1/\mu} = \left(\frac{3}{2} + \gamma\right)\left(n^{1+1/\mu}\right)^{2-1/(\mu+1)}, \quad n \to \infty.$$

Следовательно, учитывая равенство $\lambda = \mu + 1$, находим

$$\lim_{n\to\infty} \frac{\left(n^{1+1/\mu}\right)^{2-1/\lambda}}{\mathcal{E}(n)} = \left(\frac{3}{2} + \gamma\right)^{-1}.$$

Отсюда и вытекает справедливость следствия.

## Литература

*Владимир Григорьевич Найденко*
*Юрий Леонидович Орлович*

*Институт математики НАН Беларуси*
*ул. Сурганова, 11,*
*220072 Минск, Республика Беларусь*
*E-mail: naidenko@im.bas-net.by, orlovich@im.bas-net.by*